\documentclass[twocolumn,pre,amsmath,amssymb,showpacs,superscriptaddress,floatfix]{revtex4-1}

\usepackage{bm}
\usepackage{epsfig}
\usepackage[usenames]{color}
\usepackage{graphicx}

\newcommand{\diag}{\mathrm{ diag}}

\DeclareMathOperator{\Imag}{Im}
\DeclareMathOperator{\Intt}{Int}
\begin{document}

\author{N.~M.~Chtchelkatchev}
\affiliation{Institute for High Pressure Physics, Russian Academy of Sciences, Troitsk 142190, Moscow Region, Russia}
\affiliation{Department of Theoretical Physics, Moscow Institute of Physics and Technology, 141700 Moscow, Russia}
\affiliation{L.D. Landau Institute for Theoretical Physics, Russian Academy of Sciences, Moscow 117940, Russia}

\author{A.~Glatz}
\affiliation{Materials Science Division, Argonne National Laboratory, Argonne, Illinois 60439, USA}

\date{\today}

\title{Analytical Description of the Quantum-Mesoscopic-Classical Transition in Systems with Quasi-Discrete Environment}

\begin{abstract}
We investigate dynamic properties of inhomogeneous nano-materials, which  appear in analytical descriptions typically as series of $\delta$-functions with corresponding Gibbs weights. We focus on observables relevant for transport theories of Josephson junction arrays and granular systems near the superconductor -- insulator transition. Furthermore, our description applies to the theory of tunnel junctions exchanging energy with a ``bath'', the latter having a discrete spectrum.
Using the matrix theta-function formalism we find an analytical expression for the transport characteristics capturing the complete temperature driven transition from the quantum to the classical regime.
\end{abstract}

\maketitle

\section{Introduction\label{Sec:Intro}}

The investigation of observables having singular or highly oscillating behaviors at microscopic scales, which become smooth after appropriate coarse graining is a typical occurrence in many research areas of modern physics. A detailed theoretical investigation and description is therefore highly desirable and important. Quasiclassical Green's functions have been introduced in the transport theory of spatially nonuniform superconductors and superconducting hybrid structures as an ''envelop'' approximation for exact Green's functions quickly oscillating at Fermi wavelength scales, see Ref.~\cite{Golubov_rmp} for a review. One result of this coarse-gaining of the order parameter is the Landau-Khalatnikov theory of second order phase transitions~\cite{Landau_theory}. Envelop approximations of highly oscillating solutions of hydrodynamic equations  gave an opportunity to achieve analytical and numerical advances in hydrodynamics~\cite{Zakharov}. In the study of granular superconducting materials, Josephson junction arrays (JJa) and strongly disordered superconducting films near the superconductor -- insulator transition, the current-voltage characteristics and the dynamic conductivity show singular behaviors described by weighted superpositions of delta-functions~\cite{Efetov,FVB,strange_papers,SEA2009,Fazio_review,Beloborodov07,condmat2011}. The imaginary part of the two-particle Green's function in ultrasmall metallic granules, where the electron spectrum is essentially discrete, have the form of a series of delta-functions. The transport characteristics of ultrasmall tunnel junctions exchanging energy with a quantum bath with discrete spectrum have a similar singular behavior~\cite{Ingold1991,Ingold-Nazarov,nazarov_Blanter_book}. An important and in general unsolved problem is the construction of the envelop approximation for an observable (which is typically measured) in case the theoretically derived expression is a countable superposition of delta-functions. Remarkably, this problem has a general analytical solution going beyond standard approximations that allows transforming the discrete series into integrals. We found  the envelop approximation for singular observables analytically, using the Jacoby theta-function formalisms~\cite{Theta_matrix}. Our calculations can be  physically interpreted using the language of the Landau-Hopf turbulence~\cite{Landau,Hopf,condmat2011}, at least qualitatively.

In this paper we derive the envelop approximation for the real-valued function $I(\omega)$ with real argument $\omega$:
\begin{gather}\label{eq:I}
I(\omega)=-\sum_{\{\mathbf n\}}\Imag\frac{\pi^{-1}P_{\mathbf n}}{\omega-\omega_\mathbf{n}+i0}=\sum_{\{\mathbf n\}}P_{\mathbf n}\,\delta(\omega-\omega_\mathbf{n}).
\end{gather}
Here, $\mathbf n$, is a $N$-dimensional vector of quantum numbers, $\omega_{\mathbf n}$ is a scalar real function of $\mathbf n$ and the summation is performed over all configurations of $\mathbf n$ with weights (probabilities) $P_{\mathbf n}\geq 0$. In physical applications $\omega$ is usually a variable with dimension of energy, $\omega_{\mathbf n}$ is closely related to (the difference of) the energy levels of a quantum system, while $P_{\mathbf n}$ is the Gibbs distribution probability. In the most general situation when all components of $\mathbf n$ belong to a countable set, standard approximations like the Euler-Maclaurin asymptotic~\cite{Euler-Maclaurin} do not work anymore. In that case one can not naively integrate out the $\delta$-functions in \eqref{eq:I}. Our main result deals with this case. We find the conditions when the density of the $\delta$-functions starts to increase similar to the development of Landau-Hopf chaos and find the envelope approximation for $I(\omega)$ in that case. We discuss the use of our results for solutions of specific physical problems mentioned above.

The case with linear $\omega_{\mathbf n}$ and $P_{\mathbf n}$ with bilinear exponent is one of the most relevant for applications. We focus on the situation when
\begin{gather}\label{eq:Ip}
\omega_{\mathbf n}=\mathbf{e}\cdot\mathbf{n},\qquad P_{\mathbf n}=\exp\left(-\frac{\mathbf{n}^\tau\cdot E\cdot\mathbf{n}}2\right).
\end{gather}
Here $\mathbf n=(n_1,n_2,\ldots,n_N)^\tau$ is a vector of integer numbers, $\mathbf{e}$ is a (fixed) vector and $E$ is a symmetric positively defined $N\times N$ matrix. We will also show that our investigation is not crucially dependent on the specific form of $P_{\mathbf n}$ in~\eqref{eq:Ip}.

The case where $P_{\mathbf n}$  decays quickly with $|\mathbf n|$ (all eigenvalues $\lambda_i$ of the matrix $E$ are much larger than unity) is the ``quantum'' limit. Then the set of numbers $\mathbf n$ is essentially discrete. Moreover $P_{\mathbf n}$ effectively restricts $\mathbf n$ to a subset with small $|\mathbf n|$. In this case the graph of $I(\omega)$ looks like a sparse sequence of isolated $\delta$-function peaks.

Rather straight-forward is the opposite ``classical'' case where $P_{\mathbf n}$  depends only slightly on $\mathbf n$ (all  eigenvalues $\lambda_i$ of the matrix $E$ are much smaller than unity and $N$ is sufficiently large). Then the Euler-Maclaurin approximation is applicable  and one can replace the discrete sum over $\mathbf n$ in Eq.~\eqref{eq:I} by an integration over $d\mathbf n$. The result is $I(\omega)\propto \exp(-\omega^2p/2)$, where $p$ is constant.

The most interesting case is the ``mesoscopic'' case when the eigenvalues $\lambda_i$ of the matrix $E$ are slightly smaller or of the order of unity and $N$ is not too large. This case is in between the quantum  and the classical case.  Then the restriction for the choice of $\mathbf n$ is rather weak and there are a many $\mathbf n$-vectors solving the inequality
\begin{equation}\label{eq:ineq}
    |\omega-\mathbf{e}\cdot\mathbf{n}|<\sigma,
\end{equation}
where $\sigma\ll 1$ is the width of the $\delta$-function. [$\delta$-functions in physical applications have always some small width, $\sigma$, due to, e.g., interaction with a heat bath (dissipation).] If the components of the vector $\mathbf e$ are integer numbers (commensurate) then only if $\omega$ is close to an integer [the fractional part of $\omega$ is smaller than $\sigma$] there is vector $\mathbf n$ -- the solution of inequality~\eqref{eq:ineq}. However, in case the components of vector $\mathbf e$ are not commensurate -- as it is most natural in physical systems -- the formal solution of Eq.~\eqref{eq:ineq} for arbitrary $\omega$ is only applicable in the classical regime where no (integer) restriction of $|\mathbf n|$ exists. In the mesoscopic regime, $\mathbf n^\tau E\mathbf n\leq 1$, the set of $\omega$ for which Eq.\eqref{eq:ineq} has a solution is restricted. But the effective measure of this set is much larger then in the case when $\mathbf e$ has commensurable components. This implies that the effective density of $\delta$-functions in Eq.~\eqref{eq:I} strongly increases in a given $\omega$ interval, but is still discrete. We will further refer to this observation as the $\delta$-function ``condensation''. As the result the singular part of $I(\omega)$ becomes relatively small and $I(\omega)$ has a smooth envelop approximation, $\propto\exp(-\omega^2p/2)\vartheta(\omega)$. We emphasize that the functional behavior of $I(\omega)$ is different from the classical limit due to the nontrivial factor $\vartheta(\omega)$.  We find that $\vartheta$ is the  one-dimensional Jacoby theta-function, $\vartheta(\omega,\tau^*)$, where the real parameter $\tau^*>0$ depends on $E$ and $\mathbf{e}$ and qualitatively shows to what degree the components of $\mathbf e$ are incommensurate. We explicitly calculate $\tau^*$, which behaves in the classical limit as $\tau^*\to\infty$ resulting in $\vartheta\to 1$.

In order to investigate $I$ in the regime where the $\delta$-functions condensate, it is quite ineffective to use the $\delta$-function representation of $I$, Eq.~\eqref{eq:I}, directly. Instead, we rewrite $I$ in terms of matrix $\Theta$-function, $\Theta(z,\mathcal T)$ [generalizing the approach suggested in Ref.~\cite{condmat2011}], where the real symmetric matrix $\mathcal T$ has one zero eigenvalue, while the other eigenvalues are positive. This zero mode is the manifestation of $\delta$-functions in Eq.~\eqref{eq:I}. We investigate the class of the $\Theta$-functions with the (nearly)degenerate $\mathcal T$; it is an interesting problem itself. Using the matrix $\Theta$-function representation of $I$ we can analytically and numerically investigate $I(\omega)$ in the regime where the $\delta$-function condensation takes place. In a limiting case we reproduce the results of Ref.~\cite{condmat2011}. We relate the nature of the strong increase of the $\delta$-function density to the chaotic behavior of  quasi-periodic functions. Our results can help understanding  the transport theory of Josephson junction arrays and the superconductor-insulator transition. Finally, we discuss how stable our results related to the $\delta$-function condensation in Eq.~\eqref{eq:I} are with respect to the choice of the shape of the weight functions $P_{\mathbf n}$ other than given in Eq.~\eqref{eq:I}.

The structure of our paper is the following: In Sec.~\ref{Sec:theta}  we show how problem \eqref{eq:I} can be reformulated in terms of matrix theta-functions; In Sec.~\ref{sec2} we investigate the properties of the matrix theta-function in the mesoscopic regime and in particular formulate the conditions for the delta-function condensation; In Appendix~\ref{sec3} we give a numerical receipt for an efficient calculation of the matrix theta-function; finally in the discussion section~\ref{disc}, we show that the problem we address in this paper has a number of important physical applications.

\section{ Generalized matrix theta-function \label{Sec:theta}}

We rewrite the sum in Eqs.~\eqref{eq:I}-\eqref{eq:Ip} using the Poisson's formula for summation
\begin{equation}\label{Poisson}
\sum_{n=-\infty}^{\infty}f(n)=\sum_{m=-\infty}^{\infty}\int_{-\infty}^{\infty}
f(x)e^{2\pi im x} dx\,,
\end{equation}
where $f$ is a continuous integrable  function. Function $I$ in Eq.~\eqref{eq:I} depends on the vector $\mathbf n=(n_{1},n_{2},\ldots,n_{N})^\tau$. Therefore, we have to introduce also $\mathbf m=(m_{1},\ldots,m_{N})^\tau$ and $\mathbf x$ accordingly.  Approximating the $\delta(z)$-function by a smeared Gaussian function $e^{-z^2/2\sigma^2}/\sqrt{2\pi\sigma^2}$ with $\sigma\to 0$ we get after the integration over $\mathbf x$:
\begin{widetext}
\begin{gather}\label{eq:I_G}
  I(\omega)=\sum_{\mathbf m}\frac1{\sqrt{2\pi\sigma^2(2\pi)^N\det G}}
\exp\left\{ -\frac{\omega^2}{2\sigma^2}+\left(-(2\pi)^2 m_im_j+i\frac{2\pi \omega}{\sigma^2} [e_im_j+e_jm_i]+\left[\frac{\omega}{\sigma^2}\right]^2e_ie_j\right)\frac 12\langle x_i x_j\rangle\right\}.
\end{gather}
\end{widetext}

Here
\begin{eqnarray}
    G_{ij}=E_{ij}+e_ie_j/\gamma,\qquad \gamma=\sigma^2,
    \\
   \text{and}\,\, \langle x_i x_j\rangle=[G^{-1}]_{ij}.
\end{eqnarray}

\subsection{$\sigma$-expansion}
We simplify Eq.~\eqref{eq:I_G} by expanding it over $\sigma$. Then we find,
\begin{multline}\label{eq:I_G_sigma0}
I(\omega)=\sum_{\mathbf m}\frac{\exp(-\omega^2p/2)}{\sqrt{(2\pi)^{N+1} g}}\times
\\
\exp\left( -2\pi^2 m_i\langle x_i x_j\rangle m_j+i2\pi \omega \mathbf a\cdot\mathbf m\right),
\end{multline}
where the explicit analytical expressions for $g=\lim_{\sigma\to 0}\gamma\det G$, $p=-\lim_{\sigma\to 0} \gamma^{-1}[\gamma^{-1}\mathbf{e}^\tau\cdot G^{-1}\cdot \mathbf{e}-1]$ and $\mathbf{a}=\lim_{\sigma\to 0}\gamma^{-1}G^{-1}\mathbf{e}$ can be found in Appendix, Eqs.~\eqref{identity_g}-\eqref{eq_K}.

Introducing the Riemann theta-function~\cite{Abramovitz,Theta_matrix},
\begin{equation}\label{eq:theta}
\Theta(\omega,\mathcal T)=\sum_{\mathbf m} e^{i2\pi \omega \mathbf m\cdot \mathbf a-\pi \mathbf m^\tau\mathcal T\mathbf m},
\end{equation}
we rewrite $I$ as follows:
\begin{eqnarray}\label{Itheta}
    I(\omega)&=&\frac{\exp(-\omega^2p/2)}{\sqrt{(2\pi)^{N+1} g}}\Theta(\omega,\mathcal T),
    \\\label{Tau}
    \mathcal T&=&2\pi K.
\end{eqnarray}
The proof that $\mathcal T\mathbf e=0$ is given in Appendix, Eq.~\eqref{G_dot_e}. Therefore, the $\Theta$-function matrix parameter is degenerate resulting in the appearance of the $\delta$-functions in Eq.~\eqref{eq:I}. It is worth noting that  $\mathbf a$ is generally not an eigenvector of $\mathcal T$ (such that $\mathcal T\mathbf a\neq 0$).

An important role play the eigenvalues $\tau_i$ of matrix $\mathcal T$. We will distinguish the cases when $\tau_i$ are smaller or larger than unity below. Without loss of the generality we can assume that the eigenvalue corresponding to the eigenvector $\mathbf e$ has the index one: $\tau_1=0$. Then for the other eigenvalues follows: $\tau_{i>1}>0$.

We kept $\sigma=0$ above which resulted in $\tau_1=0$. In typical physical realizations, the $\delta$-function always have a finite width $\sigma$ as was mentioned in the introduction. If we preserve the sub-leading terms in $\sigma>0$ in $I$ then $\tau_1\sim \sigma^2\ll \tau_{i>1}$. Perturbations of the other parameters in Eq.~\eqref{eq:theta} produce sub-leading corrections to the shape of $I(\omega)$.


\begin{figure}[t]
  \center
  \includegraphics[width=0.9\columnwidth]{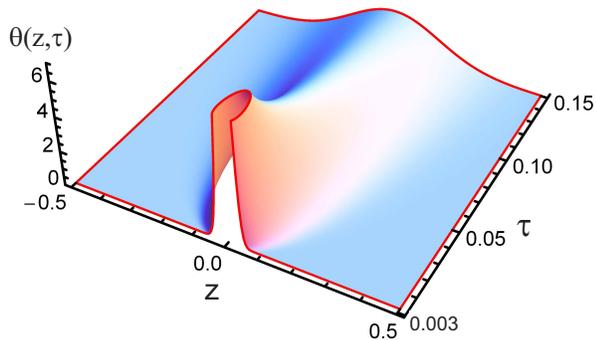}\\
  \caption{(Color online) The graph of $\vartheta(z,\tau)$.}\label{fig_theta3d}
\end{figure}
\subsection{The ``minimal model''.\label{minmodel}}

There are special cases when $\mathbf a=\mathbf e$ and the parameters $p$, $g$, and $\mathcal T$ have very simple forms. One case corresponds to a diagonal matrix $E$ as follows from the proof given in Appendix~\ref{ApB}. The case, when $E$ has the additional structure: $E_{ij}=\beta e_i\delta_{ij}$, where $\beta$ is a parameter and $\beta e_i>0$, is important for transport physics in JJAs and dirty superconducting films near the superconductor-insulator transition~\cite{FVB,condmat2011}.  We will refer to this case as to the ``minimal model''. It should be noted that $\mathbf e$ in this case is not an eigenvector of $E$ in general. This case is of theoretical interest since it is possible to find $G^{-1}$ exactly and $\det G$ analytically in all orders of $\gamma$:
\begin{gather}
    [G^{-1}]_{ij}\beta=-\frac{1}{\beta\gamma+\sum_{k}e_k}+\frac{\delta_{ij}}{e_i},
    \\\label{eq:detg}
    \det G=\beta^N\frac{(\beta\gamma+\sum_i e_i)\prod_k e_k}{\beta\gamma}.
\end{gather}
Then we get provided $\sum_ie_i\neq 0$:
\begin{equation}\label{Tau_e}
   \mathcal T_{ij}=\frac{2\pi }{\beta} \left(\frac{\delta_{ij}}{ e_i}-\frac1{\sum_k e_k}\right).
\end{equation}
Similarly we find using Eq.~\eqref{eGieb2} that
\begin{equation}\label{p2}
   p= \frac\beta{\sum_{i} e_{i}}.
\end{equation}
Expressions Eqs.~\eqref{eq:detg}-\eqref{p2} generalize similar results in Ref.~\cite{condmat2011} obtained for the limit where $\beta>0$, $\gamma\to 0$, $\sum_k e_k=N$, and $\mathbf e$ is close to the vector $(1,1,\ldots,1)^\tau$.

We took $\sigma=0$ in Eq.~\eqref{Tau_e} such that $\tau_1=0$. If we take into account $\sigma\ll 1$ then we should correct all matrix elements, $\mathcal T_{ij}$, by an additional term, $ 2\pi\sigma^2 /[(\sum_i e_i)^2]$  corresponding to $\tau_1=2\pi\sigma^2/|\mathbf e|^2$.

\section{Matrix $\Theta$-function in the mesoscopic regime \label{sec2}}

Below we study how the properties of the function $\Theta$ in Eq.~\eqref{Itheta} depend on the parameters $\mathbf e$, $E$, and especially $N$.

\subsection{$N=1$} We start our analysis with the simplest case, $N=1$. Then we get from Eq.~\eqref{eq:theta} that $\Theta$ reduces to the usual Jacobi theta-function:
\begin{equation}\label{eq:theta1}
\vartheta(\omega,\tau)=\sum_{m} e^{i2\pi \omega m-\pi \tau m^2}.
\end{equation}
For $\tau\gg 1$, $\vartheta(\omega,\tau)\to 1$ and for $\tau\ll 1$, $\vartheta(\omega,\tau_1)\approx\sum_{n} \delta(\omega-n)$, see Fig.~\ref{fig_theta3d} for an illustration. The case $N=1$ is marginal to some extent because then $\tau=\tau_1\sim\sigma^2$ and so $\vartheta$ is always a set of $\delta$-functions; formally there is no classical limit for $N=1$. In the following we focus on the case $N>1$ only.

\begin{figure}[tb]
  \center
  \includegraphics[width=\columnwidth]{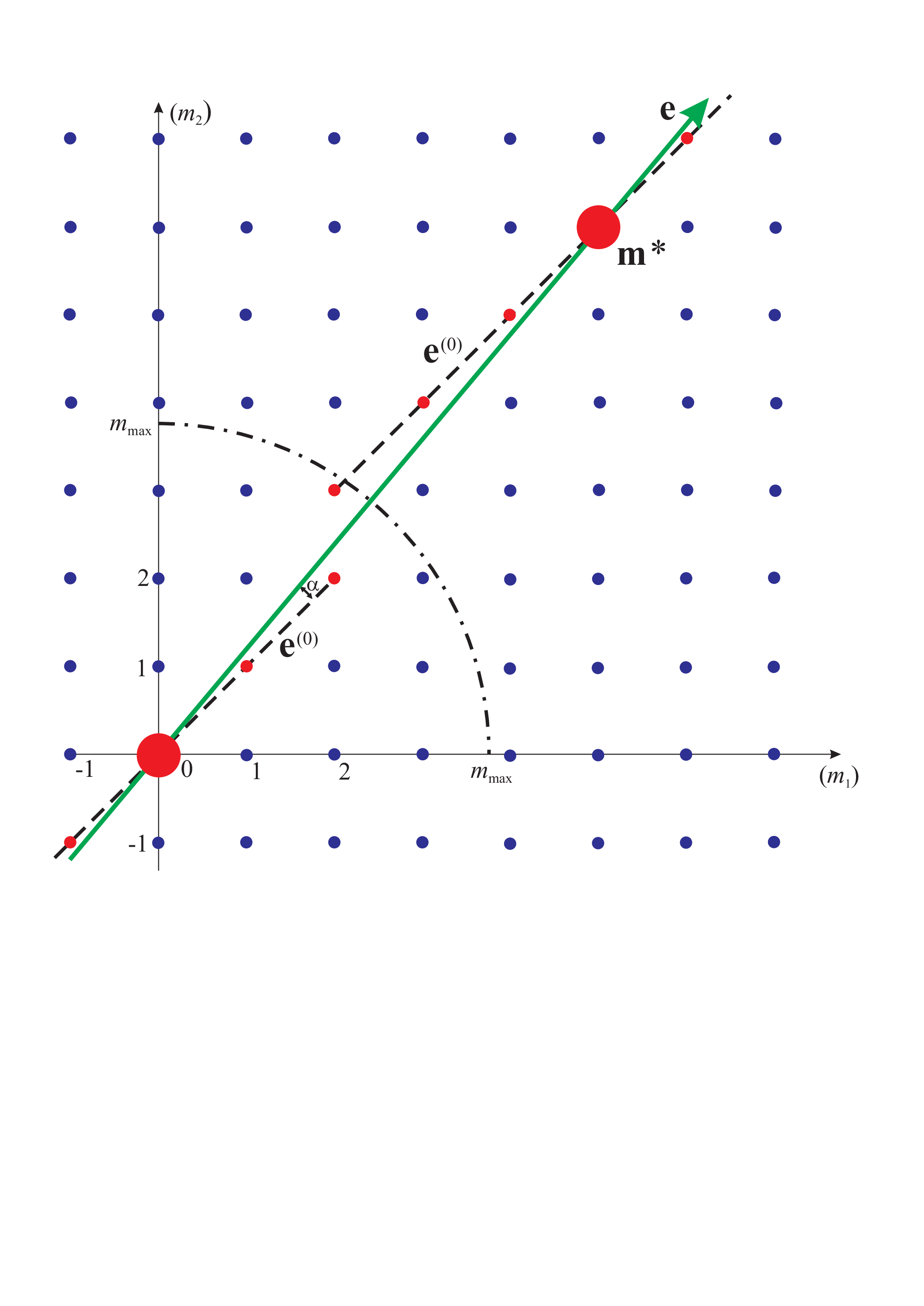}\\
  \caption{(Color online) The dots schematically show the set $\mathbf m$ for $N=2$. Without loss of the generality we assume that $\mathbf e$ (direction singled out by the green line) is close to $\mathbf e^{(0)}=(1,1)$. The dotted lines are parallel to $\mathbf e^{(0)}$. The red points belong to the subset of $\mathbf m$ that gives the leading contribution to $\Theta$ in the mesoscopic regime, Eq.~\eqref{eq:theta}.}\label{condition}
\end{figure}
\subsection{$N>1$}

The classical limit as follows from Eq.~\eqref{Poisson} can be found by setting $\mathbf m=0$ in Eqs.~\eqref{eq:I_G_sigma0}-\eqref{Itheta}. Then $\Theta(\omega,\mathcal T)\to 1$ and as follows from Eq.~\eqref{Itheta}
\begin{equation}\label{eqIcll}
    I(\omega)\to \frac{\exp(-\omega^2p/2)}{\sqrt{(2\pi)^{N+1} g}}.
\end{equation}
We recall that in the mesoscopic regime $\lambda_i\ll 1$ ($\tau_{i>1}\gg 1$). In the mesoscopic (quantum) regime $I(\omega)$ strongly differs from Eq.~\eqref{eqIcll} because the prefactor, $\Theta(\omega,\mathcal T)$, in Eq.~\eqref{Itheta} behaves in a non-trivial way.

The definition of the matrix $\Theta$-function, Eq.~\eqref{eq:theta}, includes the sum over $\mathbf m$. The components of $\mathbf m$ are integer numbers. So  $\mathbf m$ can be treated as edge vectors of the nodes of an effective cubic crystal in a $N$-dimensional space, see Fig.~\ref{condition}. It follows from the definition of the $\Theta$-function, Eq.~\eqref{eq:theta}, that the leading contribution to $\Theta$ gives $\mathbf m$ nearly collinear to $\mathbf e$.

We focus first on the case when $\mathbf e$ is close to the crystallographic axis $\mathbf e^{(0)}$ of the effective crystal. Then the subset of $\mathbf m $, $\mathbf m^{(0)}=m \mathbf e^{(0)}$, where $m=0,\pm1,\ldots$, gives the leading contribution to $\Theta$:
\begin{multline}\label{Theta_to_theta}
    \Theta(\omega,\mathcal T)\approx \sum_{\mathbf m} e^{i2\pi \omega m\mathbf e^{(0)}\cdot \mathbf a-\pi m^2 (\mathbf e^{(0)})^\tau\mathcal T\mathbf e^{(0)}}=
    \\
    \vartheta(\omega \mathbf e^{(0)}\cdot \mathbf a,\tau^*),
\end{multline}
and
\begin{gather}\label{eq:tau_star}
    \tau^*= (\mathbf e^{(0)})^\tau\mathcal T\mathbf e^{(0)}=\delta \mathbf e^\tau\mathcal T\delta \mathbf e,
\end{gather}
where $\delta \mathbf e=\mathbf e-\mathbf e^{(0)}$. The vectors $\mathbf e$ and $\mathbf e^{(0)}$ should be close such that the
following condition is fulfilled:
\begin{equation}\label{eq:cond_tau1}
\pi m_{\rm max}^2\tau_1> 1,
\end{equation}
which ensures a complete delta-function overlap (we recall that the $\delta$-function width is $\propto \sigma$, while $\tau_1\propto \sigma^2$). Here the integer $m_{\rm max}=\Intt[1/|\mathbf e-\mathbf e^{(0)}|]$, where $\Intt$ is the integer part. This condition indicates how far from the origin the distance between the points on the lines O$\mathbf e$ and O$\mathbf e^{(0)}$ becomes of the order of the effective crystal period, see Fig.~\ref{condition}.

\begin{figure}[tb]
  \center
  \includegraphics[width=\columnwidth]{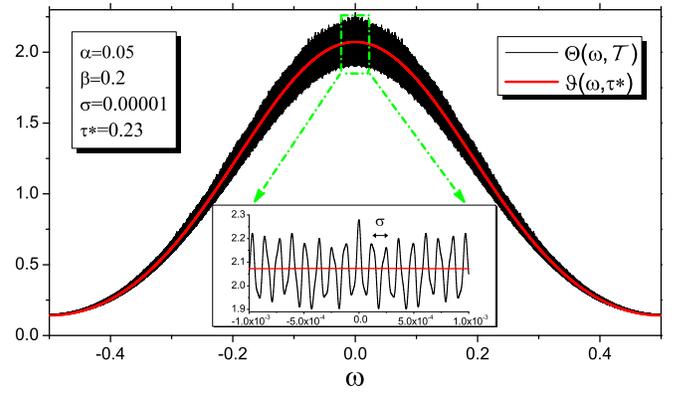}\\
  \caption{(Color online) The graph of $\Theta(\omega,\mathcal T)$ for $N=8$ [black color]. The minimal model has been used. Taking the vector from the set of the gaussian random numbers with zero average and the dispersion $\alpha$ we generated $\delta\mathbf e=\mathbf{e}-\mathbf{e}^{(0)}$, where $\mathbf{e}^{(0)}=(1,\ldots,1)^\tau$. The red line corresponds to $\vartheta(\omega,\tau^*)$ -- the envelop function~\eqref{eq:Imain}. The inset shows that the $\delta$-function width $\sigma$ is the minimal characteristic frequency of the matrix $\Theta$-function.}\label{fig_theta}
\end{figure}
Expression~\eqref{Theta_to_theta} is in fact the envelop approximation for the matrix theta-function. The $\Theta$-function quickly oscillates as the function of $\omega$ on the scale of the order of $\sigma$ on top of the envelop function~\eqref{Theta_to_theta}, see Fig.~\ref{fig_theta}. The amplitude of the oscillations does not exceed $\exp(-\pi \tau_1 (m_{\rm max})^2)$. It is worth noting that condition~\eqref{eq:cond_tau1} is sufficient but not necessary. Even beyond the limitations set by  condition~\eqref{eq:cond_tau1} the envelop function Eq.~\eqref{Theta_to_theta} usually still  approximates $\Theta$ quite well. This case is illustrated in Fig.~\ref{fig_theta}. Numerical calculations~\cite{condmat2011} show that the envelope of the $\Theta$-function follows well Eq.~\eqref{Theta_to_theta} while $|\mathbf e-\mathbf e^{(0)}| \lesssim \sigma e^{A N\ln N}$, where $A$ is a constant of  order unity. This exponential factor is closely related to the density of the $\delta$-functions in Eq.~\eqref{eq:I}, which have in the mesoscopic regime weights of the same order.

If $\tau^*\lesssim \sigma^2$ then $\vartheta(\omega \mathbf e^{(0)}\cdot \mathbf a,\tau^*)$ reduces to a superposition of $\delta$-functions, see Eq.~\eqref{eq:theta1} and Fig.~\ref{fig_theta3d}. These small values of $\tau^*$ appear when $\mathbf e=\mathbf e^{(0)}$, see Eq.~\eqref{eq:tau_star}. In case the components of $\mathbf e$ are integer numbers (they are commensurable) then the $\delta$-function singularities of $I(\omega)$ remain in the mesoscopic regime as well. However, if $\mathbf e$ and $\mathbf e^{(0)}$ are even slightly different, $\tau^*$ can easily become of order unity and $\vartheta$ becomes a smooth function of $\omega$.  Returning to Eq.~\eqref{eq:I} we will have a strong increase of the $\delta$-function density and their final overlap, when the components of $\mathbf e$ become not commensurable.

\begin{figure}[tb]
  \center
 \includegraphics[width=\columnwidth]{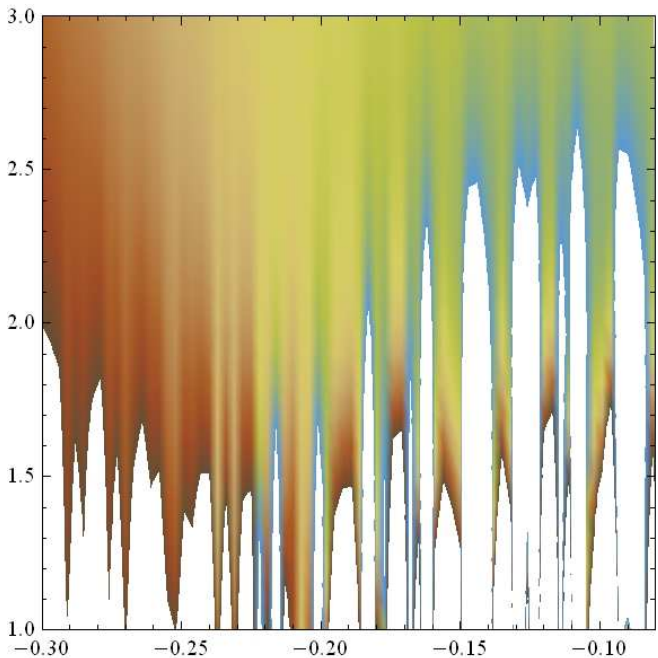}\\
  \caption{(Color online) Density graph of $\Theta(\omega,\mathcal T)$ for $N=8$. The axes $OX$ corresponds to $\omega$ while $OY$ axes corresponds to $\beta^{-1}$. When $\beta>1$ the theta function behaves as the discrete set of $\delta$-peaks. With decreasing $\beta$, more peaks become visible, for $\beta\approx 1$ the peaks start merging and for $\beta \ll 1$ the matrix theta function becomes nearly smooth.}\label{fig3d}
\end{figure}

To conclude this section, we emphasize that Eq.~\eqref{Theta_to_theta} is one of our main results. It shows that in the mesoscopic regime $I(\omega)$ is nearly continuous (plus a relatively small and quickly oscillating background) and the form of the function $I(\omega)$ is nontrivial. We find the envelop approximation of $I(\omega)$ analytically:
\begin{equation}\label{eq:Imain}
    I(\omega)\backsimeq \vartheta(\omega \mathbf e^{(0)}\cdot \mathbf a,\tau^*)\, \frac{\exp(-\omega^2p/2)}{\sqrt{(2\pi)^{N+1} g}}.
\end{equation}
The accuracy of the prefactor in Eq.~\eqref{eq:Imain} is illustrated in Fig.~\ref{fig_theta}. The density graph~\ref{fig3d} shows the evolution of the matrix $\Theta$-function in the minimal model when the parameter $\beta$ switches the model from the quantum to the mesoscopic regime. The numerical calculations are briefly described in Appendix~\ref{sec3}.

\section{Discussion \label{disc}}
\subsection{Time representation}

The time representation helps to understand the properties of $I(\omega)$ from another point of view.  For a special case it was shown in \cite{condmat2011} that the Landau-Hopf turbulence scenario can account for the merging of the $\delta$-functions in  Eq.~\eqref{eq:I} in the mesoscopic regime. Here we apply the ideas developed in Ref.~\cite{condmat2011} for the general case. To this end we express $I(\omega)$, Eq.~\eqref{eq:I}, in time representation, $I(t)=\int I(\omega) e^{-i\omega t} d\omega$, and find:
\begin{eqnarray}\label{eq:It}
    I(t)=\sum_{\mathbf n} P_{\mathbf n} e^{i \sum_{i=1}^N n_i \varphi_i(t)},
    \\ \label{eq:phi}
    \varphi_i(t)=\omega_i t,\qquad \omega_i=e_i.
\end{eqnarray}

\begin{figure}[tb]
  \center
 \includegraphics[width=0.7\columnwidth]{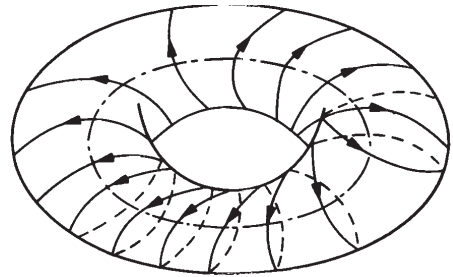}\\
  \caption{A graphical representation of  Eqs.~\eqref{eq:It}-\eqref{eq:phi} describing an  open winding path on a two-dimensional torus surface.}\label{fig3}
\end{figure}

Function $I(t)$ belongs to the set of \textit{quasi-periodic} functions \cite{QPF_def}. If we take $\varphi_i$ as coordinates in an $N$-dimensional space then the trajectory $\varphi_i(t)$ is modeled by a curve on a torus $T$ that wraps around without ever exactly coming back on itself if the $\omega_i$ are \textit{incommensurate}, see Fig.~\ref{fig3} for $N=2$. The path covers the torus surface densely everywhere. If we return to Eq.~\eqref{eq:I} then this property would mean that the $\delta$-function positions are densely distributed.   In the quantum regime only $\mathbf n$ with $|\mathbf n|\leq 1$ contribute to $I(t)$ such that not more than one frequency appears in the exponent in Eq.~\eqref{eq:It}. The classical limit formally corresponds to a torus with an infinite number of dimensions.

Our consideration for the condensation of the $\delta$-functions in Eq.~\eqref{eq:I} can be extended to the more general case of arbitrary weights $P_{\mathbf n}>0$ which decay quickly with $\mathbf n$~\cite{chtch}. In this case, the time representation of $I$ would consist of quasi-periodic functions as well and the topological argument of the path covering a torus surface densely would be applicable again. In this more general case $I(\omega)$ cannot be reduced to the matrix theta-function. However, the Poisson transformation of $I$ with an appropriate $\sigma$-expansion still allows to reduce $I$ to an ``easy to handle'' form for analytical and numerical investigation in the regime when the $\delta$-functions start to merge.

It is worth noting that expressions similar to Eq.~\eqref{eq:It} appear in many applications. For example, it describes the velocity field $\mathbf v(\mathbf r)$ of a liquid when turbulence develops according to the Landau-Hopf scenario~\cite{Landau,Hopf,condmat2011}, $\mathbf{v}(t)=\sum \mathbf A_{p_1p_2\ldots p_N} \exp\left\{i\sum_{i=1}^N p_i\,\varphi_i(t)\right\}$. Here the developed turbulence corresponds to the case of large $N$ \cite{footnote2}.  Quasi-periodic functions describe quasi-periodic motion of mechanical systems~\cite{landau1} and often appear in the theory of differential equations, see e.g., Ref.~\cite{Novikov}.

\subsection{Superconductor-insulator transition in granular systems and JJAs}

In the study of granular superconducting materials and Josephson junction arrays (JJA) transport properties are of central importance.
The most interesting case corresponds to the quantum limit when the Josephson coupling $E_J$ between the granules is much smaller than the characteristic Coulomb energy $E_c$ of the Cooper pair located on a granule. The transport problem was solved within linear response approaches, see e.g.,  \cite{Efetov,FVB,strange_papers,SEA2009,Fazio_review,Beloborodov07}. It was shown that the conductivity $\sigma_{\scriptscriptstyle JJA}$ of the granular superconductor is a singular function in the leading order over $E_J$. It behaves as the countable superposition of $\delta$-functions with the Gibbs weights.

For example, the conductivity in a JJA according to Ref.~\cite{Efetov} has the  following form [which can also be derived from Eq.\eqref{eq:I} in appropriate limiting cases]:
\begin{equation}\label{eq:sigma_efetov}
\sigma_{\scriptscriptstyle JJA}(\omega)\sim \sum_{\vec n}\delta(\omega-b_in_i)\exp\left(-\frac{(e^*)^2 n_i B_{ij}n_j}{2T}\right),
\end{equation}
where $\omega$ is the frequency, $T$ is the temperature, the indices $i$, $j$ label the islands of the JJA, $B$ is the inverse capacitance matrix of the JJA and $b_i=e^*(B_{1i}-B_{2i})$. Here $e^*=2e$ is the charge of the Cooper pair and $e$ is the electron charge [we will use the system of units where $e=k_B=\hbar=c=1$]. The integer numbers $n_i$ show the effective number of Cooper pairs sitting on the island $i$. We assume here and below the Einstein sum convention. The exponential weights in Eq.~\eqref{eq:sigma_efetov} come from the Gibbs distribution. The delta-functions ensure energy conservation during the processes of Cooper pair tunneling from one granule of the JJA to its neighbor. They cannot be easily integrated out in Eq.~\eqref{eq:sigma_efetov} since changes of the Coulomb energy are essentially discrete because of charge quantization within granules.

The size of the JJA arrays can be quite small, especially in the case of one-dimensional (1D) arrays. Therefore, the number of junctions $N$ in the JJA is typically finite, $N\gtrsim 10$. On the other hand the interaction matrix $B_{ij}$ can be rather quickly decaying with $|i-j|$. Generally, we cannot use standard statistical physics approaches based on the thermodynamic limit $N\to \infty$ trying to smear out the $\delta$-function singularities in the observables, as it was done in Eq.~\eqref{eq:sigma_efetov}. An important question is to understand the nature of the $\delta$-function singularities in the observables and finding systematic ways of their regularization.

One way to overcome the difficulties related to the $\delta$-functions in transport observables of the JJA was proposed in Ref.~\cite{SEA2009}. That calculation was based on the assumption of an energy band for Cooper-pair tunneling. However, this band can form in JJa with nearly identical granules but should be suppressed by the disorder in typical disordered JJAs~\cite{condmat2011}.

In Refs.~\cite{FVB,strange_papers,condmat2011} an attempt was made to find the transport characteristics in the disordered JJAs. The model of Refs.~\cite{FVB,strange_papers} leads to the conductivity behaving according to Eq.~\eqref{eq:I} with $E_{ij}\propto E_c e_i\delta_{ij}$ and $\mathbf e$ close to the vector $(1,1,\ldots,1)^\tau$ (a limiting case of our minimal model). It was shown in Ref.~\cite{condmat2011} that although $\mathbf e$ has non-commensurate components, the delta-functions merge and the conductivity may become a smooth function of its parameters. In this paper we do not restrict the choice of $E$ and $\mathbf e$ like in Ref.~\cite{condmat2011}.

\subsection{Effective dimension of the $\Theta$-function}


The properties of the matrix-$\Theta$-function strongly depend on the number $d$ of the nonzero components of $\mathbf a$.  This number can be treated as an effective dimension of the $\Theta$-function. Generally $1\leq d\leq N$. Without loss of the generality we can assume that the first $d$ components of $\mathbf a$ are nonzero while the others take zero values. We can apply a reduction procedure if $d<N$: to take the sum over $m_{i}$ with $i>d$ in Eq.~\eqref{eq:theta}. We finally obtain a function very similar to the matrix-theta function but now with a summation over $d$-dimensional integer vectors. For example, $d=2<N$ for the problem considered in Ref.~\cite{Efetov} and $d=N$ for Ref.~\cite{FVB}.

\subsection{Discrete environment spectra}
The problem of the evolution of a quantum system interacting with a (quantum) environment of (soft) modes is being studied for more than 60 years, but more important to be solve than ever, see e.g., Refs.~\cite{Fok,Davydov,Nazarov1989,Devoret_basic,Girvin1990,Averin1990,Ingold1991,Ingold-Nazarov,chtchprl2008}. Usually it is implied that the environment has a continues spectrum of modes. That condition is important since it typically avoids the the appearance of the $\delta$-functions like in Eq.~\eqref{eq:I} in observables of the quantum system. A discrete environment, on the other hand,  cannot simply absorb arbitrary amounts of energy, but rather only discrete energy portions in quantums on the order of its level spacing.  In other words, the set of quantum modes can serve as a ``bath''  only when its spectrum is continuous~\cite{Fok}, which we will clarify below.
\begin{figure}[tb]
  \center
 \includegraphics[width=\columnwidth]{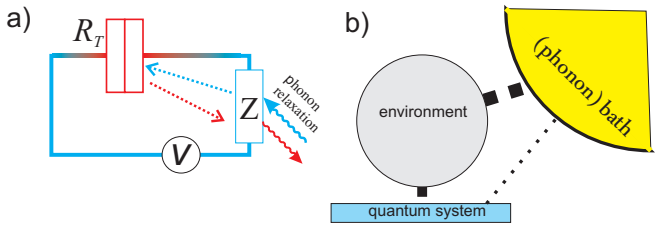}\\
  \caption{ A tunnel junction interacting with the environment of electromagnetic or many-body excitations sketched as effective impedance $Z$. The wavy lines show schematically the exchange of the energy between the junction and the environment~\cite{chtchprl2008}.  b) The environment is not necessary described by the equilibrium density matrix~\cite{chtchprl2008}. Here we sketched the environment interacting with a thermal bath and the quantum system.  }\label{figeh}
\end{figure}

An important problem where a quantum system interacts with an environment, is the problem of quantum transport through an ultrasmall tunnel junction, see Fig.~\ref{figeh}, and Ref.~\cite{Ingold-Nazarov} for a review. If the contacts of the junction are superconducting, the supercurrent is given by
\begin{equation}\label{Isv}
    I_s(V)=\pi E_J^2\left(P(2V) - P(-2V)\right),
\end{equation}
where $P(\omega)$ is the probability to exchange energy $\omega$ with the environment and $E_J$ is the Josephson energy of the junction.

Next, we concentrate on the shape of $P$. According to Ref.~\cite{Ingold-Nazarov}
\begin{equation}
    P(\omega)=\frac 1 {2\pi}\int dt \langle\exp[ie^*\hat\phi(t)]\exp[-ie^*\hat\phi(0)]\rangle e^{i\omega t},
\end{equation}
where $\hat \phi$ is the ``charge transfer'' operator, $e^*=2$ for Cooper pairs and $\langle\ldots\rangle$ denotes averaging over the density matrix $\hat \rho$ of the environment. [We recall that we have chosen units where $k_B=\hbar=e=c=1$]. Since the environment is isolated, it can be described by the Hamiltonian $H_{\rm env}$ with a set of energies $\epsilon_\alpha$ and eigenfunctions $|\alpha\rangle$. In general we have $\hat \rho=w_{\alpha\beta}|\alpha\rangle\langle\beta|$. Writing the phase operator in Heisenberg representation explicitly, we get $\exp[ie^*\hat\phi(t)]=\hat U^{-1}(t) \exp[ie^*\hat\phi(0)]U(t)$, where $\hat U(t)=\exp(-i\hat H_{\rm env} t)$. This way we find:
\begin{multline}\label{eq:Pd}
P(\omega)=\sum_{\alpha\beta\gamma\eta\lambda} \delta(E_\alpha-E_\beta+E_\gamma-E_\eta+\omega)\times
\\
w_{\lambda\alpha}\left(e^{ie^*\hat\phi(0)}\right)_{\beta \gamma}\left(e^{-ie^*\hat\phi(0)}\right)_{\eta\lambda}.
\end{multline}
Thus $P(\omega)$ is represented as the superposition of $\delta$-functions. Only if the spectrum of the environment is continuous we can introduce a continuous density of states and integrate out the $\delta$-functions in Eq.~\eqref{eq:Pd}.

As long as the spectrum of the environment is discrete, we can  assume without loss of generality that the Greek indices labeling the levels are a set of integer numbers. If the environment is represented by oscillator modes (or quantum rotators), $E_\alpha-E_\beta+E_\gamma-E_\eta+\omega$ is linear in the indices labeling the environment states like in Eq.~\eqref{eq:I}. If, in addition, the environment is in local equilibrium, such that $w_{\lambda\alpha}\propto \delta_{\lambda,\alpha}\exp(-E_\alpha/T_{\rm env})$, where $T_{\rm env}$ is the environment temperature, the structure of Eq.~\eqref{Isv} is the same as the structure of Eq.~\eqref{eq:I}.

To complete our investigation we focus on the quasi-particle current $J_q$ through the tunnel junction shown in Fig.~\ref{figeh}.  According to Refs.~\cite{Nazarov1989,Devoret_basic,Girvin1990,Averin1990,Ingold1991,Ingold-Nazarov,chtchprl2008} we have
          \begin{equation}\label{eq:current}
                  J_q=\left(\overrightarrow{\Gamma} - \overleftarrow{\Gamma}\right)\, ,
           \end{equation}
where $\overrightarrow{\Gamma}$ ($\overleftarrow{\Gamma}$) is the tunneling rate from the left (right) to the right (left), and, for a single junction,
   \begin{equation}\label{eq:S_begin_total}
        \overrightarrow{\Gamma}=\frac1{R_{\scriptscriptstyle {\mathrm T}}}
        \int_{\epsilon\epsilon'}f_\epsilon^{(1)}
        (1-f_{\epsilon'}^{(2)})P(\epsilon-\epsilon')\, ,
   \end{equation}
where $f^{(1,2)}$ are the electronic distribution functions within the left (right) electrodes, and $R_{\scriptscriptstyle{\mathrm T}}$ is the bare tunnel resistance, representing the interaction of electrons with the bath. Here $P(\omega)$ is the probability for electron quasi-particle to lose the energy $\omega$ to the environment; it is given by Eq.~\eqref{eq:Pd} with $e^*=1$. The backward scattering rate is given by $\overleftarrow{\Gamma}\propto\int_{\epsilon\epsilon'}f_\epsilon^{(2)} (1-f_{\epsilon'}^{(1)})P(\epsilon-\epsilon')$. [If the contacts are superconducting, we can account for that by introducing the quasi-particle densities of the states in the contacts \cite{Ingold-Nazarov} in Eq.~\eqref{eq:S_begin_total}. Eq.~\eqref{eq:current} would then give the quasi-particle current.] If the environment is absent and the relaxation is provided by a phonon bath,  $P(\epsilon)=\delta(\epsilon)$ and  Eq.~\eqref{eq:S_begin_total} reproduces the conventional Ohm law. It follows from  Eqs.~\eqref{eq:current}-\eqref{eq:S_begin_total} that the integration over the energy  removes the $\delta$-functions of the environment and the quasiparticle current is not as singular as the supercurrent when the environment has a discrete spectrum of modes.

To summarize, the problem we solve in this paper is closely related to the problem of a discrete environment exchanging energy with a quantum conductor. It should be emphasized that the environment should not be necessary located somewhere outside the quantum conductor. On the contrary, it can be part of the conductor itself. Such a situation is realized in JJAs, see e.g., Ref.~\cite{Efetov}, or in very dirty conductors where the transition between discrete and continuous ``built-in'' environments is closely related to many-body localization, Refs.~\cite{Gornyi2005,BAA2006,chtchprl2008}.

\section{Conclusions}

We have shown that  observables having a form as described by Eq.~\eqref{eq:I} can be rewritten in terms of the matrix theta-function. We demonstrated that the $\delta$-functions typically condensate in the mesoscopic regime such that the observable $I$ we focus on becomes a nearly continuous function. We found the envelop function for $I$ analytically and therefore, our results can help to understand and describe transport properties in a number of strongly correlated systems.

\section{Acknowledgments}
The authors thank A.~Petkovic and V.~Vinokur for active discussions in the initial stage of the work; we also thank M.~Fistul for helpful comments and T. Baturina for interest to our work.

This work was supported in part by the Russian Foundation for Basic Research (Grant No. 11-02-00341) and by the U.S. Department of Energy Office of Science under the Contract No. DE-AC02-06CH11357.

\appendix

\section{Parameters of the $\Theta$-function. Analytical expressions. \label{ap:A}}

\subsubsection{Identity 1}
The matrix $E$ can be diagonalized by the orthogonal transformation $U$, such that $U^\tau \lambda U=E$, where $\lambda=\diag(\lambda_1,\ldots,\lambda_N)$ is the diagonal matrix of the eigenvalues. Then expanding the determinant of $G$ over $\gamma$ we get the following relation:
\begin{equation}\label{identity_g}
g=\lim_{\sigma\to 0}\gamma\det G=\det E\,\sum_{i} \frac{(\tilde e_{i})^2}{\lambda_{i}},
\end{equation}
where $\mathbf{\tilde e}=U\cdot\mathbf e$.

\subsubsection{Identities 2,3}
Similarly we can prove, using an induction procedure, that
\begin{equation}\label{eGie}
    \lim_{\sigma\to 0} \gamma^{-1} e_i(G^{-1})_{ij}e_j=1,
\end{equation}
and
\begin{multline}\label{eGieb2}
    p=-\lim_{\sigma\to 0} \gamma^{-1}[\gamma^{-1}e_i(G^{-1})_{ij}e_j-1]=
    \\
    \frac{\det E}{g}=\left(\sum_{i} \frac{(\tilde e_{i})^2}{\lambda_{i}}\right)^{-1}. 
\end{multline}
It should be noted that $p>0$.

\subsubsection{Identity 4}
\begin{equation}\label{gammaG_dot_e}
    \lim_{\sigma\to 0}\gamma^{-1}(G^{-1})_{ij}e_j=a_i.
\end{equation}
We will see that the vector $\mathbf a$ is the characteristic direction of the $\vartheta$-function. It is worth noting that generally $\mathbf e$ and $\mathbf a$ are not parallel in Euclidean space. The vector $\mathbf a$ can be found explicitly:
\begin{equation}\label{eq:a}
    \mathbf a=g^{-1}\det(E)\, U^\tau \left(
                         \begin{array}{c}
                           \frac{\tilde e_1}{\lambda_1}  \\
                           \frac{\tilde e_2}{\lambda_2} \\
                           \ldots \\
                           \frac{\tilde e_N}{\lambda_N} \\
                         \end{array}
                       \right).
\end{equation}

\subsubsection{Identity 5}
It follows from Eq.~\eqref{gammaG_dot_e} that
\begin{equation}\label{G_dot_e}
    \lim_{\sigma\to 0}(G^{-1})_{ij}e_j=0.
\end{equation}

\subsubsection{Identity 6 }
Finally, we find the inverse of the matrix $G$ at $\sigma\to 0$. This is somewhat tricky since $G$ becomes singular while, following Eq.~\eqref{identity_g}, $G^{-1}$ becomes degenerate. Nevertheless there is a finite nontrivial limit for $G^{-1}$ at $\sigma\to 0$:
\begin{multline}\label{eq_K}
  K=\lim_{\sigma\to 0}G^{-1}=g^{-1}\det(E)\, U^\tau\cdot
  \\
   \left(
      \begin{array}{cccc}
        k_{1} & -\frac{\tilde e_1\tilde e_2}{\lambda_{1}\lambda_2} & \ldots & -\frac{\tilde e_1\tilde e_N}{\lambda_{1}\lambda_N} \\
        -\frac{\tilde e_2\tilde e_1}{\lambda_2\lambda_{1}} & k_{2} & \ldots & -\frac{\tilde e_2\tilde e_N}{\lambda_{2}\lambda_N} \\
         \ldots & \dots & \ldots & \ldots \\
        -\frac{\tilde e_N\tilde e_1}{\lambda_N\lambda_{1}} & -\frac{\tilde e_N\tilde e_2}{\lambda_N\lambda_{2}} & \ldots & k_{N} \\
      \end{array}
    \right)\cdot U
\end{multline}
The diagonal elements have the structure:
\begin{equation}
   k_{i}=\sum_{j\neq i}\frac{\tilde e_j^2}{ \lambda_j},
\end{equation}
where $i=1,\ldots,N$.

\section{Diagonal $E$-matrix\label{ApB}}

We assume that $\mathbf e$ is an eigenvector of  matrix $E$ corresponding -- without loss of generality -- to eigenvalue $\lambda_1$. Then $\tilde e=|\mathbf e|(1,0,\ldots,0)^\tau$, and therefore
\begin{eqnarray}
  g&=&\frac{|e|^2}{\lambda_1}\det E  \\
  K&=&U^\tau\diag(0,\lambda_2^{-1},\lambda_3^{-1},\ldots,\lambda_N^{-1})U,
\end{eqnarray}
where $\lambda_{i>1}=2\pi/\tau_i$.

Most important in this case is
\begin{equation}
\mathbf a=\mathbf e.
\end{equation}
This identity can be proven in a diagonal representation of $E$. Without loss of the generality $\mathbf{\tilde e}$ corresponds to the eigenvalue $\lambda_1$ of $E$. Then we can take again $ \mathbf{\tilde e}=|\mathbf e|(1,0,\ldots,0)^\tau$ and
\begin{gather}
  U G^{-1}U^\tau=\diag\left(\frac{1}{\lambda_1+\gamma^{-1}},\frac 1{\lambda_2},\ldots,\frac 1{\lambda_N}\right)\quad \Rightarrow
  \\
   \mathbf{\tilde a}\equiv\lim_{\sigma\to 0}\gamma^{-1}U G^{-1}U^\tau \mathbf{\tilde e}=\mathbf{\tilde e}.
\end{gather}
So, $\mathbf a=U^\tau \mathbf{\tilde a}=U^\tau \mathbf{\tilde e}=\mathbf e$.

\section{$I(\omega)$, numerical investigation \label{sec3}}

An important task is to verify  the accuracy of Eq.~\eqref{eq:Imain} and the envelope approximation of the matrix $\Theta$-function numerically. The required calculations of the matrix $\Theta$-function for $N>1$ requires care since one should sum up many quickly oscillating functions. We have found an effective numerical algorithm explicitly relying on the existence of the soft mode $\mathbf e$ of the matrix $\mathcal T$.

\begin{figure}[tb]
  \center
  \includegraphics[width=\columnwidth]{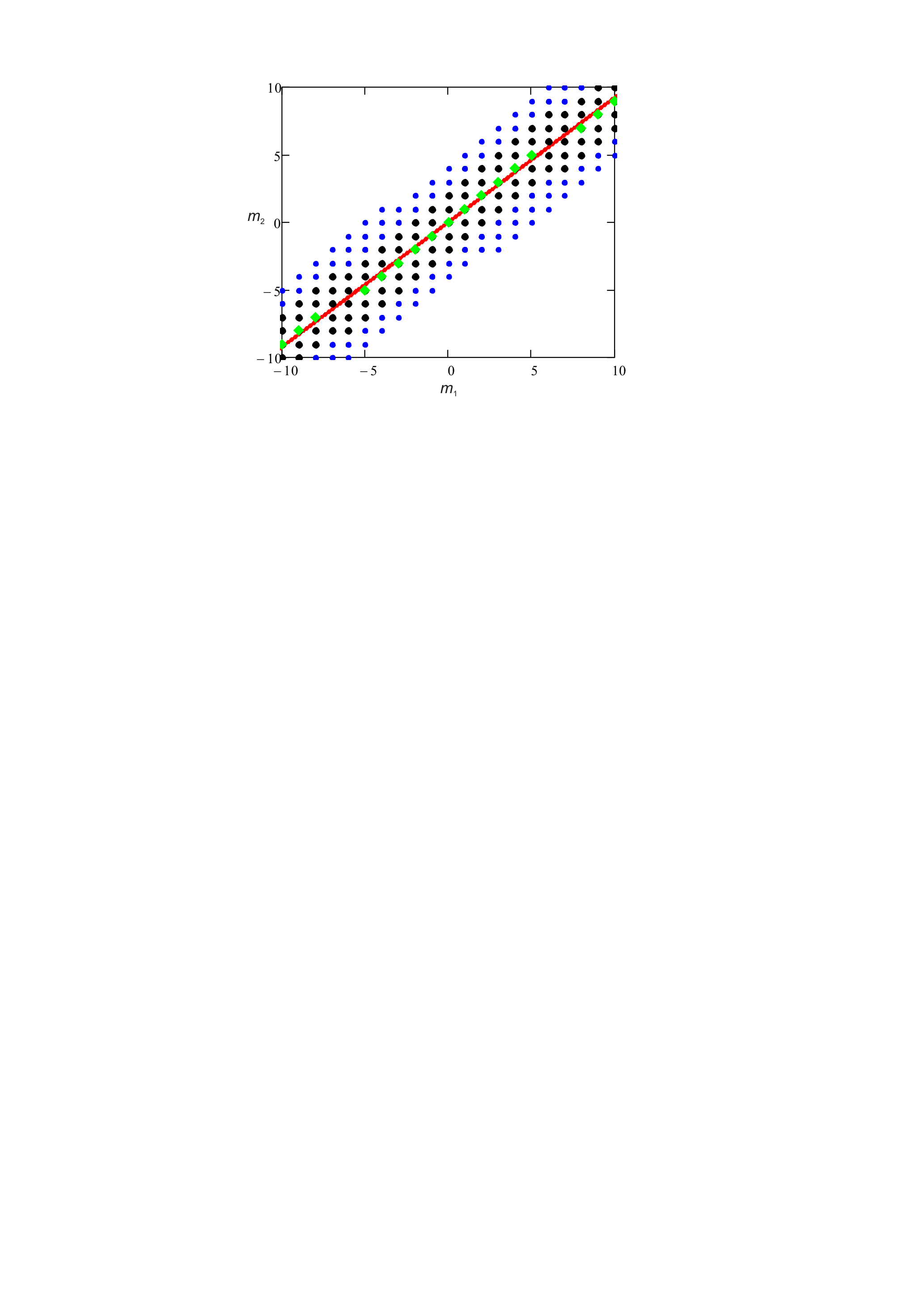}\\
  \caption{(Color online) The sketch of the set of points $\mathbf m$, in Eq.~\eqref{eq:theta}, that belong to the ``cylinder'' volume and satisfy the condition $\mathbf m^\tau \mathcal T\mathbf m\lesssim1$. Here the solid line is directed along $\mathbf e=(1.04,0.96)^\tau$. The green diamonds distinguish the points of the effective crystal that give the leading contribution to $\Theta$ for $\tau_2\lesssim10$; the black points should be added if $\tau_2\lesssim 2$ while the small blue points become important for $\tau_2<1$. }\label{fig_tube}
\end{figure}

The matrix $\mathcal T$ can be expressed explicitly through its eigenvalues $\tau_1,\ldots,\tau_N$ and the corresponding eigenvectors $\mathbf e$ and $\mathbf h^{(i)}$, $i=2,\ldots,N$ [$\mathbf e\perp \mathbf h^{(i)}$]
\begin{equation}
    \mathcal T=\tau_1 \mathbf e\cdot\mathbf e^\tau+\sum_{i=2}^N\tau_i \mathbf h^{(i)}\cdot(\mathbf h^{(i)})^\tau.
\end{equation}
The set of  $\mathbf m$, see Eq.~\eqref{eq:theta}, form a $N$-dimensional cubic crystal. We construct a cylinder in this space with  generating lines $\mathbf e$ and an elliptical support with main directions $\mathbf h^{(i)}$. Doing numerical calculations we take into account only points of the crystal, $\mathbf m$, in Eq.~\eqref{eq:theta} that belong to the cylinder volume, see Fig.~\ref{fig_tube}. The proportions of the cylinder depend on the calculation  accuracy. The ratio of the cylinder height to its characteristic diameter is of the order of $\min\{\tau_{i>1}\}/\tau_1\gg 1$. The case $\tau_{i>1}\gg 1$ is the most interesting since then the $\delta$-functions in Eq.~\eqref{eq:I} are expected to condensate. Then only $\mathbf m$, the nearest neighbors to the line directed along $\mathbf e$, should be taken into account, see green diamonds in Fig.~\ref{fig_tube} and red dots in Fig.~\ref{condition}. This property strongly reduces the numerical efforts compared to a direct calculation of $I(v)$ using Eq.~\eqref{eq:I}.

A typical graph of $\Theta(\omega,\mathcal T)$ for $N=8$ is shown in Fig.~\ref{fig_theta} [black line], the red line shows the envelop function represented according to Eq.~\eqref{Theta_to_theta} by the one dimensional (Jacoby) $\Theta$-function. The inset shows that the $\delta$-function width $\sigma$ is the minimal characteristic ``frequency'' of the matrix $\Theta$-function. It follows also that the matrix $\Theta$-function can be approximated by the one-dimensional (Jacoby) $\Theta$-function with specially chosen parameters. For producing the graph~\ref{fig_theta} we have used $\mathcal T$ from Eq.~\eqref{Tau_e}. This minimal model we have chosen for two reasons: 1) $\mathcal T$ is parameterized by a minimal set of parameters, 2) this model has some relation to physical applications, see Refs.~\cite{FVB},\cite{condmat2011}.

The vector $\mathbf e$ we expressed as $\mathbf e=\mathbf e^{(0)}+\delta \mathbf e$, where  we chose $\mathbf e^{(0)}=(1,1,\ldots,1)$ according to Ref.~\cite{condmat2011}. The components of $\delta \mathbf e$ are generated by a Gaussian random number generator with zero average and the variance $\alpha$. For simplicity we used the restriction, $\sum_i \delta \mathbf e_i=0$. Then we get $\tau^*=2\pi\left[\frac{\sigma^2}{4}+ \alpha^2 \right]$ that agrees with Ref.~~\cite{condmat2011}.

If $\mathbf e=\mathbf e^{(0)}$, the components of $\mathbf e$ are commensurate and $\Theta(\omega)$ behaves just as the superposition of $\delta$ functions shifted by a constant period of the order unity along the $z$-axis as follows from Eq.~\eqref{eq:I}. But if $\delta \mathbf e\neq 0$ then the components of $\mathbf e$ are not commensurate, there is no periodicity in the $\delta$-function set, as it is illustrated in Fig.~\ref{fig_theta} for $\alpha=0.05$.

We implied nearly everywhere above that $\beta<1$. If this is not the case and $\beta>1$, $P_n$ in Eq.~\eqref{eq:I} quickly decay with growing $|\mathbf n|$ and we can disregard all $\mathbf n$ with $|\mathbf n|>1$ within acceptable error bars. It is clear that in that case $I(\omega)$ [as well as $\vartheta$] behaves always as a set of $\delta$-functions with some finite distance from each other. Disorder in $\mathbf e$ only slightly shifts the positions of the $\delta$-functions on the $\omega$-axis and there is no $\delta$-function condensation. What happens when we go from $\beta<1$ to $\beta>1$ is illustrated in Fig.~\ref{fig3d}. When $\beta>1$ the theta function behaves as a discrete set of the $\delta$-peaks. With decreasing $\beta$ more peaks develop, for $\beta\approx 1$ the peaks start merging and for $\beta \ll 1$ the matrix theta-function becomes nearly smooth.


\begin{thebibliography}{99}
\bibitem{QPF_def} The function $f (t)$ is the quasi-periodic function with periods $w_1,w_2,\ldots,w_n$ if $f (t) \equiv F (t, t, ..., t)$ where $F(t_1, t_2,\ldots,t_n)$ is periodic function of $t_1,t_2,\ldots,t_n$ with the periods, $w_1,w_2,\ldots,w_n$, respectively. All the periods $w_1,w_2,\ldots,w_n$ are strictly positive, and their reciprocals $p_1,p_2,\ldots,p_n$ are rationally linearly independent. The quasiperiodic function can be represented by the series: $f(t)=\sum c_{k_1,\ldots,k_n}\exp(i t\sum_ik_ip_i)$.
\bibitem{footnote2} In real fluids the Landau-Hopf turbulence scenario typically does not realize since it does not take into account properly the nonlinear effects~\cite{Kolmogorov,Turbulence}.
\end{thebibliography}
\end{document}